\documentclass[a4letter
]{IEEEtran}
\IEEEoverridecommandlockouts
\usepackage{cite}
\usepackage{amsmath,amssymb,amsfonts}
\usepackage{algorithmic}
\usepackage{graphicx}
\usepackage{textcomp}
\usepackage{xcolor}
\usepackage{mathtools}
\usepackage{xfrac}
\usepackage{nicefrac}
\def\BibTeX{{\rm B\kern-.05em{\sc i\kern-.025em b}\kern-.08em
    T\kern-.1667em\lower.7ex\hbox{E}\kern-.125emX}}
 \usepackage[all]{background}
 \usepackage{stackengine}
 \setstackEOL{\\}
 \setstackgap{L}{\normalbaselineskip}
 
  \SetBgContents{\color{gray}{\tiny \Longstack{IFIP/IEEE VLSI-SoC 2024 - Accepted Version; DOI:10.1109/VLSI-SoC62099.2024.10767817}}}
 \SetBgPosition{4.5cm,1cm}
 \SetBgOpacity{1.0}
 \SetBgAngle{0}
 \SetBgScale{1.8}

\usepackage{todonotes}

\usepackage{hyperref}
\usepackage[defaultcolor=red, draft]{changes}

\usepackage{comment}

\begin{document}

\title{Secure Software/Hardware Hybrid In-Field Testing for System-on-Chip
\thanks{\IEEEauthorrefmark{1} These authors contributed equally to this work. \\
This work has been supported by DAIS (https://dais-project.eu/), which has received funding from the ECSEL Joint Undertaking (JU) under grant agreement No 101007273. The JU receives support from the European Union's Horizon 2020 research and innovation program and Sweden, Spain, Portugal, Belgium, Germany, Slovenia, Czech Republic, Netherlands, Denmark, Norway, and Turkey. }

}


\makeatletter 
\newcommand{\linebreakand}{%
  \end{@IEEEauthorhalign}
  \hfill\mbox{}\par
  \mbox{}\hfill\begin{@IEEEauthorhalign}
}
\makeatother 

\author{\IEEEauthorblockN{Saleh Mulhem\IEEEauthorrefmark{1}, Christian Ewert\IEEEauthorrefmark{1}, Andrija Ne\v{s}kovi\'{c}, Amrit Sharma Poudel,\\ Christoph Hübner, Mladen Berekovic, and Rainer Buchty}\\
\IEEEauthorblockA{\textit{Institute of Computer Engineering, Universität zu Lübeck, Lübeck, Germany} \\
\{name.surname\}@uni-luebeck.de}
}

%




\maketitle

\begin{abstract}
Modern Systems-on-Chip (SoCs) incorporate built-in self-test (BIST) modules deeply integrated into the device's intellectual property (IP) blocks.
Such modules handle hardware faults and defects during device operation.
As such, BIST results potentially reveal the internal structure and state of the device under test (DUT) and hence open attack vectors. 
So-called result compaction can overcome this vulnerability by hiding the BIST chain structure but introduces the issues of aliasing and invalid signatures. 
Software-BIST provides a flexible solution, that can tackle these issues, but suffers from limited observability and fault coverage.
In this paper, we hence introduce a low-overhead software/hardware hybrid approach that overcomes the mentioned limitations.
It relies on (a) keyed-hash message authentication code (\textsc{KMAC}) available on the SoC providing device-specific secure and valid signatures with zero aliasing and (b) the SoC processor for test scheduling hence increasing DUT availability.
The proposed approach offers both on-chip- and remote-testing capabilities.
We showcase a RISC-V-based SoC to demonstrate our approach, discussing system overhead and resulting compaction rates.
\end{abstract}
\begin{IEEEkeywords}
Secure Built-In Self-Test, System-on-Chip, KMAC, In-Field Testing 
\end{IEEEkeywords}
\section{Introduction}
\label{sec:introduction}

The increasing complexity and functionality of System-on-Chip (SoC) devices require extensive and detailed testing to ensure reliable behavior during their operation \cite{2022_ayache_dual_cone_v_model}.
For this, built-in self-test (BIST) provides a mechanism for testing without physical access to the device under test (DUT).
BIST is usually a hardware mechanism, but can be software-based (SBIST)\cite{2007_wang_soc_test_architectures_nanometer_design_testability}. It interrogates the internal device status via a scan chain, thus giving a detailed insight into the system state to be analyzed by automated test equipment (ATE). 

From a security perspective, the adversary can exploit the BIST's full observability of the DUT internal status to compromise the SoC \cite{10530017}.
Therefore, it is considered a side-channel information leakage \cite{2020_cui_secure_scan_design_obfuscation_hash}. 
This is particularly harmful for cryptographic SoCs and safety-critical systems. 
Relying on output-response analyzers (ORA) to compact the scan-chain output responses into a signature might sometimes help with hiding internal DUT data. 
However, this is not always the case for classical signature generation as it reveals sensitive data or information \cite{2019_li_scan_chain_based_attacks_countermeasures_survey}.
Another challenge of traditional methods is the generation of invalid signatures when $L < d$, where $L$ and $d$ are the bit length of the DUT output response and its corresponding signature, which in this case directly reveal internal DUT information.
Countermeasures to overcome such vulnerabilities are not only specific to the DUT, but potentially also introduce a vast overhead, making them not suitable for resource-constrained SoC \cite{2019_li_scan_chain_based_attacks_countermeasures_survey}.

Further, hardware testing interferes with the SoC's availability; it hence cannot be tested during normal system operation.
SBIST \cite{2007_wang_soc_test_architectures_nanometer_design_testability} provides a suitable solution for testing under operational conditions by only scheduling the test when the DUT is not on operational demand. 
By nature, the SBIST approach is self-contained. As such, the DUT observability is restricted to the self-testing device.
A detailed DUT analysis also requires an extensive database, making SBIST-only unsuitable for resource-constrained embedded devices.

The mentioned test methods exhibit the following shortcomings: 
\begin{description}
    \item[S1:] Traditional BIST methods are vulnerable to \textit{data mapping} and \textit{signature analysis} attacks \cite{2019_li_scan_chain_based_attacks_countermeasures_survey}. Thus, once a single device is compromised, all other in-field devices are vulnerable to the same attacks.
    \item[S2:]     Although the test methods achieve almost zero \textit{signature aliasing rate} (collision rate) when $L \geq d$, they generate invalid and insecure signatures when $L < d$.
    \item[S3:] To harden the test methods, extra hardware resources must be adapted to the specific device, introducing additional overhead. In the case of SBIST, also an extensive database is needed for detailed DUT analysis.
\end{description}

In this paper, we tackle these issues and introduce a secure hybrid software/hardware approach for in-field SoC testing. 
The proposed approach takes advantage of available hash function primitives in modern SoCs for secure signature generation and employs the SoC processor (CPU) to coordinate and schedule the test in a similar manner to SBIST. 
The main contribution of this work can be listed as follows:
\begin{itemize}
    \item We propose a new software/hardware hybrid approach as ORA for SoC in-field testing based on keyed-hash message authentication code (\textsc{KMAC}) for signature generation. This mechanism generates valid signatures even when $L < d$. Furthermore, a sufficiently large device-specific key ensures that potential attacks are limited to a dedicated device, and compromising one makes it difficult to compromise other devices.
    \item Inspired by SBIST, we deploy the SoC's processor (CPU) to coordinate the test in conjunction with \textsc{KMAC}-based ORA. 
    \item The proposed SoC test is flexible: It allows course-grained functional diagnosis via on-chip testing and a detailed DUT fault analysis as a new remote testing mechanism with a fault dictionary for detailed fault diagnosis.    
    \item We demonstrate the proposed hybrid approach featuring a RISC-V ecosystem.
\end{itemize}

To the best of our knowledge, this is the first work introducing this combination of \textsc{KMAC} and SBIST.
\section{Background \& Related Work}
\label{sec:background_related_work}

This section partially reviews state-of-the-art SoC in-field test approaches, particularly emphasizing solutions based on cryptographic primitives.   

\subsection{SoC In-field Test Approaches}
\label{subsec:bist_approaches}
BIST is a technique for performing in-field tests detecting SoC defects during device operation.
Additional scan chains are employed, applying test vectors on the DUT and collecting responses for subsequent analysis by the ATE.
During testing, the SoC is unavailable. This test-induced downtime needs hence to be minimized. Resulting, new test methods were introduced:
SBIST \cite{2007_wang_soc_test_architectures_nanometer_design_testability} provides a suitable solution to this challenge.
A self-test library (STL) is stored beside the normal execution programs in memory.
The processor executes the test program for STL-based tests and collects the responses compared to a golden reference.
This mechanism reduces SoC downtime as the SBIST is scheduled only when the corresponding component or system operation is available for testing.

To perform remote SoC in-field testing and diagnosing, the JTAG debug interface is often used to enable device access where local testing is not feasible \cite{2012_portolan_packet_based_jtag_remote_testing}.
The JTAG interface is connected to the local BIST scan chain. 
The ORA then compacts this test result into a signature which is sent to the remote tester for evaluation using a signature database.

\subsection{SoC Test Vulnerabilities and Countermeasures}
\label{subsec:test_vulnerabilities}

The provided deep SoC access for remote system diagnosis and testing via JTAG could reveal internal and sensitive information of SoC components via the BIST scan-chain \cite{2019_li_scan_chain_based_attacks_countermeasures_survey}. This is not just restricted to, e.g., structural information, but can potentially include sensitive data of, e.g., a cryptographic chip:  
An attacker can, for instance, access dedicated registers via internal scan-chain states to leak sensitive data such as cipher secret keys in a data mapping analysis attack \cite{2004_yang_scan_attack_des} \cite{2014_ali_scan_attack_soa_countermeasure_dft}. 
In signature analysis attacks, the attacker targets the compacted test response to reveal the secrets of \textsc{AES} \cite{2012_ege_diff_scan_attack_aes_x_compactor, 2011_darolt_scan_attack_countermeasure_response_compactor}, \textsc{DES} \cite{2012_kodera_scan_attack_des_scan_signatures} and RSA \cite{2012_darolt_scan_attack_rsa_industrial_countermeasures}.
In both attack scenarios, the attacker knows the internal scan-chain structure and can control the test at a fine granularity.

SoA countermeasures provide limited protection against the mentioned attacks or are rather costly \cite{2019_li_scan_chain_based_attacks_countermeasures_survey, 2018_vishwakarma_jtag_mitigation_iot_survey}.
Obfuscation techniques have been deployed to hide the internal scan-chain state and protect the DUT against data mapping analysis attacks.
Obfuscation techniques incorporate an XOR-scan and added randomization technique via physically unclonable functions (PUF) \cite{2014_banik_cryptanalysis_double_feedback_xor_chain}, state-dependent randomization \cite{2012_atobe_scan_ff_key_config_against_scan_attck_rsa}, or dynamic obfuscation provided by random number generators \cite{2018_wang_scan_test_obfuscation}.
To protect the DUT against signature analysis attacks further, encrypting \cite{2019_da_silva_prevent_scan_attack_scan_chain_encryption} or masking the interface via a PUF \cite{2018_li_puf_secure_scan_chain_design} were introduced.
It must be noted that all reviewed countermeasures require an individual solution for each possible attack.

\subsection{Cryptographic Primitives-based Solutions for SoC Test Vulnerabilities}
\label{subsec:Crypto_test}
Hash-based scan-chain obfuscation was first proposed in \cite{2020_cui_secure_scan_design_obfuscation_hash}.
A salted \textsc{SHA3-512} is utilized to hash the scan-chain output.
The salt is mainly deployed to protect the inference of small output responses from a possible rainbow attack. 
Two steps are required to generate the salt: First, a random selection of bits from the scan-chain output is applied to create a \textit{seed}, and then XORing the \textit{seed} bits results in the salt. 
Although the salt is usually a public value \cite{2010_nist_recommendation_pw_based_key_derivation}, the salt function of the salted \textsc{SHA3-512} is assumed to be confident.
A secret salt, or so-called pepper, typically requires a salt size of 112 bits \cite{2020_nist_digital_identity_guidelines}; however, it significantly slows down the hash verification.
In \cite{2012_kumar_jtag_architecture_multi_level_security}, the JTAG interface and its transmitted data were protected by AES for encrypting the communication channel.
Since the encryption of JTAG data is not required for all device life-cycles (e.g., testing, assembling, and shipping), this solution relies on privilege management to perform the encryption only when needed. 
The mentioned countermeasures for protecting the scan-chain data and communication channels introduce a non-negligible overhead to the SoC, which is unsuitable for resource-constrained devices.

In this paper, we introduce hybrid software/hardware SoC in-field testing to overcome the shortcomings of state-of-the-art solutions, particularly S1, S2, and S3 as stated in the previous section. 
\section{The Proposed Methodology}
\label{sec:methodology}

The proposed test method relies on the following pillars:
(a) the available \textsc{KMAC} on the SoC to provide device-specific secure signatures, (b) a virtual prototype (VP) of SoC to generate the golden reference signatures, and (c) the CPU to schedule the test, increase DUT availability, and support the remote testing approach.
This section covers and explains these pillars in detail. 

\subsection{Golden Reference Signature Generation}
\label{subsec:sig_generation}
The golden reference signature is generated in the early SoC design phase, 
based on reference values extracted from an SoC VP built for testing purposes at a high abstraction level \cite{2024_jaysena_directed_test_generation_survey}.
Using the hash function, signatures of (a) each component, (b) subsystems, and (c) the complete SoC VP are generated.
Fig.~\ref{fig:system_modelling_graph} shows a graph-based system modeling approach representing the complete SoC and its components.
\begin{figure}[t]
    \centering
    \includegraphics[width = .45\textwidth]{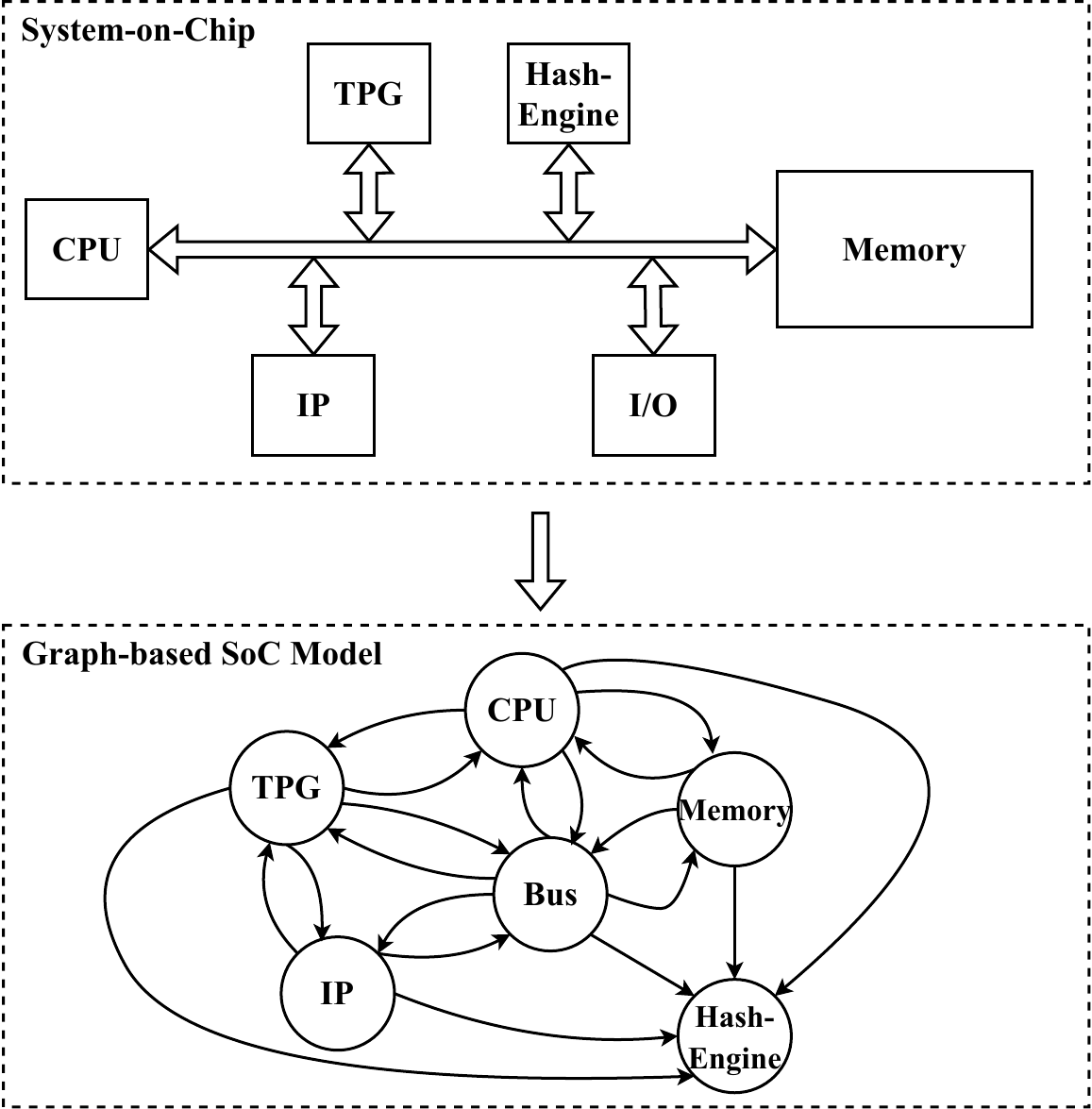}
    \caption{Signature Generation based on SoC Graph Model}
    \label{fig:system_modelling_graph}
\end{figure}
This includes the processor, memory, IPs, peripherals, and interconnection buses \cite{2008_mishra_spec_driven_test_generation_processors, 2012_grosso_sbist_system_peripherals, 2008_larsson_integrated_soc_test_framework}.
The graph-based system modeling could provide a formal and scalable approach.
Here, the interaction between the SoC and its components is a directed acyclic graph $G = (V, E)$, in which a vertex $V$ could determine an overall system state, a system component, a dedicated register and its state, or a logic gate. 
An edge $E$ could represent a system's state transition, a component, a register, a bus connecting components, or a wire connecting registers and logic gates.
The granularity of the graph-based model determines the level of detail representing the SoC on the logic level, the component functionality, or the interaction of the SoC components.

\subsection{\textsc{KMAC}-based Test Method: Setup}
\label{subsec:hash_based_test_methodSetup}
Fig. ~\ref{fig:system_test_domain} outlines the separation into test domain and DUT. 
\begin{figure}[t]
    \centering
    \includegraphics[width = .45\textwidth]{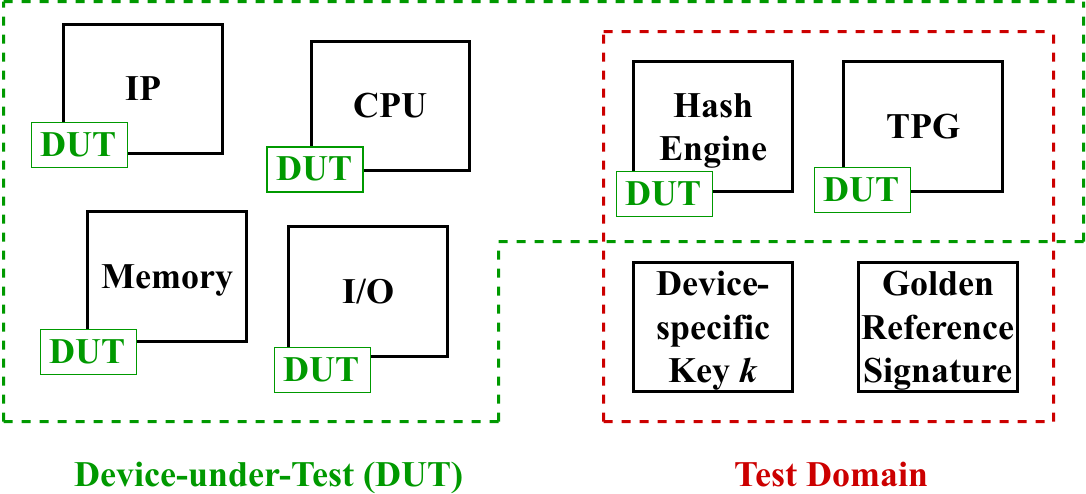}
    \caption{DUT \& Test Domain Separation}
    \label{fig:system_test_domain}
\end{figure}
It must be noted that every system component or a composition thereof, like the entire SoC, is a DUT.
This includes parts of the test domain as both \textsc{KMAC} hash engine and test pattern generator (TPG) also can be tested individually.  
These two units together with the device specific keys and the golden reference signature comprise the entity of the test domain.

\subsection{\textsc{KMAC}-based Test Method: Concept}
\label{subsec:hash_based_test_methodSetup}

\begin{figure}[t]
    \centering
    \includegraphics[width = .42\textwidth]{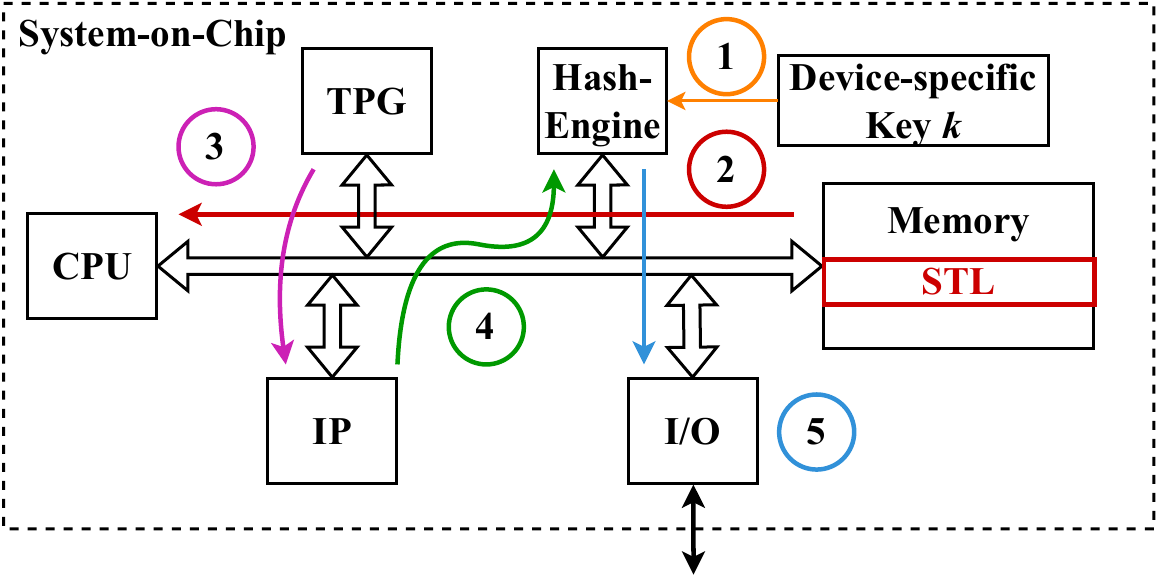}
    \caption{The proposed SoC Hash-based Test Method}
    \label{fig:soc_test}
\end{figure} 

The test mechanism is performed during normal system operation, but without disrupting it. For the proposed test methodology illustrated in Fig.~\ref{fig:soc_test}, the following steps should be carried out:

\begin{enumerate}
    \item \textsc{KMAC} is initialized with a device-specific key \textit{k}. A key manager can perform the key derivation, which is hidden from the system bus.
    \item The main CPU coordinates the 
    hardware-assisted SBIST approach using an STL. The CPU schedules the test when a component or the system is not on operational demand.
    \item The CPU requests test patterns from a linear feedback shift register (LFSR) deployed as a pseudo-random TPG and feeds them to the DUT.
    \item The output responses of the DUT are sent to the \textsc{KMAC}.
    \item Finally, \textsc{KMAC} generates the corresponding digest as a device-specific DUT signature, compared to a stored golden reference.
\end{enumerate} 
Due to the hash function's high collision resistance, numerous output responses can be fed to the hash, providing high test coverage while preserving zero-aliasing of the signature. 
In contrast to state-of-the-art in-field test approaches, our proposal provides secure testing of resource-constrained devices, where adding hardware-intensive scan chains and their countermeasures against the presented attacks is unfeasible.
If necessary, the test coverage can be further increased by extending the hybrid SBIST with low-overhead LBIST hardware \cite{2019_floridia_hybrid_online_self_test_architecture} including the hash engine for response compaction without being exploited to an external tester.

\subsection{Two Configurations of the Proposed Test Method}
Let $\mathcal{S}=\{s_1,\dots,s_n\}$ be the seeds which are deployed to generate the test vector patterns $\mathcal{V}=\{v_1,\dots,v_n\}$ by the TPG and $\mathcal{R}=\{r_1,\dots,r_n\}$ the responses by the targeted DUT, where $|r_i|=L_i$ bits for every $i=1,\dots,n$.
The set $\mathcal{H}=\{h_1,\dots,h_n\}$ represents the corresponding signatures obtained:
\begin{equation}
    KMAC^A(r_i) = H(k_A||r_i) = h_i.
    \label{eq:Kmac}
\end{equation}
$SoC^A$ indicates an SoC with a device-specific key $k_A$, and $H$ is a standard hash function with $|h_i|=d$ bits for every $i=1,\dots,n$. 
Every response signature $h_i$ determines specific diagnoses of all DUTs, such as fault-free, faulty, etc.
Therefore, we call the database including $\mathcal{S}$ and $\mathcal{H}$ the device-specific fault dictionary $D_A$ for the DUTs in $SoC^A$.
The in-field test can be proposed for two different system configurations based on the targeted applications of SoC and its design decisions.

\subsubsection{On-Chip Testing}
On-chip testing is sufficient if it ensures functional correctness of the SoC or component, reducing the database to a locally stored small device-specific fault dictionary $D_A$.
For the on-chip test, the CPU selects a seed \(s_j\) from \(\mathcal{S}\) stored in the memory and sends it to the TPG, generating the test vectors \(v_j\). The test is performed based on the method shown in Fig.~\ref{fig:soc_test}. 
The generated signature \(h'_j\) then is compared to the locally stored golden reference hash \(h_j\) to ensure the functional correctness of the DUT in $SoC^A$.

The fixed hash size enables the device to determine its complex status without requiring an excessive amount of memory.
This provides predictable test requirements from early design steps and helps integrate such in the overall design phases.

\subsubsection{Remote Testing}
The signature can be compared based on an exhaustive fault dictionary by involving an external trusted authority, enabling a detailed fault analysis.
This configuration of the proposed test method can be performed as shown in Fig.~\ref{fig:dut} and illustrated step-by-step in Table~\ref{tbl:remote_Tetsing} : 
\begin{itemize} 
    \item[(1)] The trusted remote tester sends seed \(s_j\) from \(\mathcal{S}\) to $SoC^A$. The TPG of $SoC^A$ generates test vectors \(v_j\) from \(\mathcal{V}\) based on \(s_j\).
    \item[(2)] The test vectors \(v_j\) are applied to the DUT. The corresponding response signature \(h'_j\) is generated via \textsc{KMAC} hardware engine and $SoC^A$ sends back \(h'_j\) to the remote tester. 
    Then, the DUT diagnosis can be performed by comparing the computed signature \(h'_j\) and stored golden reference \(h_j\) in the device-specific fault dictionary. 
\end{itemize}
 
\begin{figure}[t]
    \centering
    \includegraphics[width = .47\textwidth]{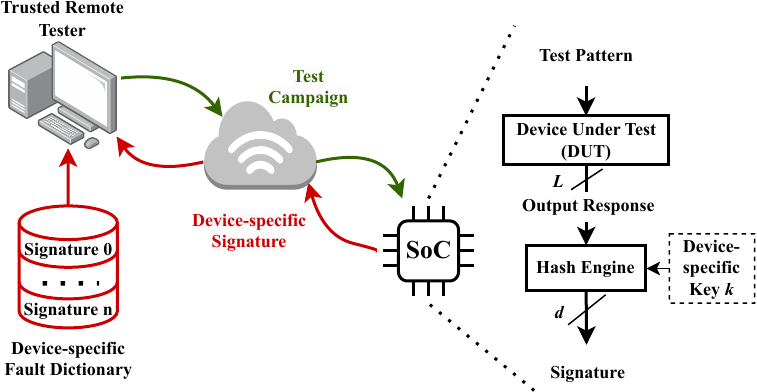}
    \caption{Proposed \textsc{KMAC}-based Test Method}
    \label{fig:dut}
\end{figure}
The remote testing leverages the test from a fast but course-grained on-chip approach testing the functional correctness of the DUT to a more sophisticated mechanism, which can determine specific faults. 
For every detectable fault, a dedicated response signature \(h_j\) is stored in the remote tester's fault dictionary enabling a detailed fault analysis of all DUTs in $SoC^A$.
\begin{table}[t]
\caption{Remote Testing Protocol}
\setlength\tabcolsep{0pt}
\def\arraystretch{1.5}
\centering
    \begin{tabular*}{\columnwidth}{@{\extracolsep{\fill}}p{.1\columnwidth}p{.35\columnwidth}p{.15\columnwidth}p{.35\columnwidth}}
    \hline
    \multicolumn{4}{c}{Device-Specific SoC Remote Testing Protocol} \\ \hline
    & \hfil Trusted Remote Tester \hfil                                &                & \hfil \(SoC^A\): DUT \hfil \\ 
    
    (1)   &  Chooses \(s_j\) from \(\mathcal{S}\).        &   \(\xrightarrow{s_j}\)  &  \(TPG(s_j)\)\(= v_j\) \newline \(DUT(v_j)\)\(= r'_j\) \newline \\
    
    (2)  & \textbf{If} \(h'_j = h_j \), the diagnose can be found based on the corresponding $r_j$ in $\mathcal{R}$   \newline \textbf{else} $h'_j$ is an invalid signature.  & \(\xleftarrow{h'_j}\)  &  \(KMAC^A(r'_j)\)\(=\) \newline \(H(k_A||r'_j)\) \(=h'_j\) \\ \hline
    \end{tabular*}
    \label{tbl:remote_Tetsing}
\end{table}
\section{Hash Selection \& Test Security Analysis}
\label{sec:signature_security_analysis}

This section introduces the selected hash function and provides a detailed security evaluation of the proposed test methodology.

\subsection{Cryptographic Hash Selection}
\label{subsec:hash_selection}
According to NIST \cite{2012_hash_recommandations_nist_fips}, hash functions ideally provide a collision resistance of $\frac{d}{2}$ with $d$ being the digest's bit length.
We chose \textsc{Keccak} \cite{2015_sha3_nist_fips} for our work. 
\textsc{Keccak} consists of a state of $b$ bits divided into a rate of $r$ bits and capacity of $c$ bits, so that $b=r+c$.
Deploying \textsc{Keccak} for ORA in BIST delivers a constant size digest as a signature for arbitrary output responses with the length $L$.
\textsc{KMAC} is \textsc{Keccak} with a device-specific key utilized to prevent signature analysis attacks in case of the attacker's knowledge about the internal test structure or rainbow table attacks when the size of the output response is small.
The signature generation, in this case, is performed by the \textsc{KMAC} \cite{2016_kmac_nist_fips} mode of \textsc{Keccak}. \textsc{KMAC} builds upon the extendable output function (XOF) \textsc{SHAKE}, which delivers two modes with different collision resistance properties:
\textsc{SHAKE128} with a collision resistance of $\min(d/2, 128)$ and \textsc{SHAKE256} of $\min(d/2, 256)$, where $d$ can be chosen arbitrary.
Thus, the digests should have a minimum length of $d_{\textsc{SHAKE128}} = 256$ and $d_{\textsc{SHAKE256}} = 512$ to provide the maximum collision resistance, respectively.
We feed the \textsc{KMAC}-\textsc{Keccak} by the DUT output response.  
The hash digest of size $d$ represents a resulting signature.
This enables using \textsc{KMAC} for complex BIST of SoC devices while providing a compact and device-specific signature with high collision resistance.

\subsection{Test Security Analysis}
The state-of-the-art ORAs are distinguished between space and time compactors.
One way of space compaction is XOR reduction \cite{2004_mitra_x_compact}, providing either a high compaction rate with low area overhead at the price of a high aliasing rate, or high area consumption when requiring lower signature aliasing.
By utilizing a serial-input signature register (SISR) or multiple-input signature register (MISR), time compaction provides a suitable solution for more significant tests while reducing hardware consumption.
We analyze our approach in comparison with SISR and MISR compaction.\\  
The aliasing probability $PA$ for SISR and MISR is calculated as follows \cite{2006_wang_vlsi_test_priciples_architectures}:

\begin{equation}
    PA_{SR}(n,L) = \frac{2^{L-n}-1}{2^{L}-1}
    \vspace{7pt}
    \label{eq:psa_sisr_misr}
\end{equation}

With this methodology, the SISR or MISR generates a signature for the DUT output response with the length $L$.
The SISR or MISR size $n$ determines the signature length $d$.
Thus, the compaction rate $CR$ of SISR and MISR is calculated as follows:

\begin{equation}
    CR_{SR}(n,L) = 1 - \frac{n}{L}
    \vspace{7pt}
    \label{eq:cr_sisr_misr}
\end{equation}

(\ref{eq:psa_sisr_misr}) \& (\ref{eq:cr_sisr_misr}) only hold when $L > n$.
A minimum DUT output response length is required to generate a valid signature.
When providing signatures with $L\leq n$, the uncompacted output response from the DUT is revealed enabling data mapping analysis attacks. 
For this, ORAs are specifically adapted to fit the output size of a DUT to provide circuit-specific signatures.
Deploying the \textsc{KMAC} mode of \textsc{Keccak} as an ORA, the aliasing probability is equal to the collision resistance based on the digest size $d$: 

\begin{equation}
    PA_{\textsc{Kmac}}(d) = \frac{1}{2^{d/2}}
    \vspace{7pt}
    \label{eq:p_keccak}
\end{equation}

The output response of the DUT is compacted to a fixed-size signature.
For this, the compaction rate of \textsc{KMAC} generated signature is calculated as follows:

\begin{equation}
    CR_{\textsc{Kmac}}(d,L) = 1 - \frac{d}{L}
    \vspace{7pt}
    \label{eq:cr_keccak}
\end{equation}

Utilizing a SISR or MISR as an ORA with equal aliasing probability to \textsc{KMAC} would require a signature register of size $n = d/2$.
For example, deploying \textsc{KMAC128} as a hash engine with a digest size $d = 256$ to match the maximum collision rate, an equal SISR/MISR-based compactor would require a register of size $n=128$, thus for all output responses $L \leq 128$ an invalid signature would be generated. 

(\ref{eq:cr_sisr_misr}) \& (\ref{eq:cr_keccak}) show the requirement for having a larger signature when deploying \textsc{KMAC} as an ORA in comparison to SISR/MISR-based signature for equal aliasing probability.
For large test responses $L\gg d$, the compaction rate of \textsc{KMAC} never the less is still sufficient, especially when performing complex SoC tests.
\begin{table}[t]
\centering
\caption{Hardware Consumption @ ZCU102 FPGA Evaluation Board}
\begin{tabular}{c|c|c|c}
                        & \textbf{LUT} & \textbf{Register} & \textbf{Slices} \\ \hline
\textbf{PULPissimo SoC} & 48639        & 42321             & 11289           \\
\textbf{KMAC128}       & 3618         & 1646              & 534             \\
\textbf{TPG}            & 59           & 71                & 26             
\end{tabular}
\label{tbl:fpga_results}
\end{table}

\section{Implementation \& Results}
\label{sec:implementation_results}

We chose a RISC-V SoC as a case study to demonstrate the proposed method.  
First, we extended the RISC-V VP \cite{8524047} with C++ \textsc{KMAC128} and TPG modules to generate the golden references. 
STL was also designed and verified in the RISC-V VP.  
Then, we integrated a \textsc{KMAC128} engine into the PULPissimo RISC-V SoC \cite{2018_schiavone_pulpissimo}.
We selected \textsc{KMAC128} and a digest size of $d = 256$~bits to match the maximum collision resistance while the hash output (signature) is kept as minimal as possible.
The signature aliasing probability in this case is $PA_{\textsc{KMAC128}}=\frac{1}{2^{128}}$, which is comparable to state of the art solutions \cite{2019_li_scan_chain_based_attacks_countermeasures_survey}; in contrast, however, this low aliasing probability persists even for the case of $L<d$.
Furthermore, we selected a key $k$ with a length of $\left|k\right|$ = 64~bits.
Thus, at least $2^{64}$ possibilities are provided to generate any signature with an arbitrary output response length of the DUT, making it a hard-to-achieve task for an attacker to perform signature analysis attacks.
To provide an integrated TPG without requiring additional memory for storing the test pattern, an LFSR with a polynomial of $32^{nd}$ degree is deployed.
The size of the TPG is adapted to match the 32-bit bus size, generating sufficient test pattern for all SoC components.
The PULPissimo RISC-V SoC with the integrated \textsc{KMAC128} and TPG as memory-mapped IPs were implemented on the Xilinx ZCU102 Evaluation Board utilizing a Zynq UltraScale+ FPGA using Vivado 2022.1.
Table~\ref{tbl:fpga_results} shows the SoC hardware small overhead for both the \textsc{KMAC128} and the TPG, indicating 7.56$\%$ of look-up-tables (LUT), 4.06$\%$ registers, and 4.98$\%$ of slices.\\
To showcase the scalability for several sizes of circuits (DUTs) and test responses, we utilized circuits from the ISCAS-85 benchmark and performed automated test pattern generation (ATPG) by the \textit{ATALANTA} tool.
The test responses were compacted into the \textsc{KMAC128}-generated signatures with a digest size of $d = 256$ bits.
To analyze signature aliasing, we performed stuck-at-fault simulations utilizing Synopsys
\textit{Z01X} for all fault locations of the benchmark circuits.
The results presented in Table~\ref{tab:iscas85_results} underline the scalability of our approach.
Our results show compaction rates up to 99.65$\%$ with no aliasing occurring after signature generation.
Even for negative compaction rates, i.e., $L<d$, valid signatures are generated without revealing circuit response data.
Thus utilizing \textsc{Kmac} for response compaction in combination with SBIST provides a powerful mechanism to locally and remotely test SoC and achieve a high fault coverage without exposing the internal BIST to a malicious tester.
\begin{table}[t]
\centering
\caption{Compaction \& Aliasing Rate Analysis utilizing ISCAS-85 Benchmark}
\resizebox{.5\textwidth}{!}{
\begin{tabular}{c|c|c|c|c|c}
\textbf{}      & \textbf{\begin{tabular}[c]{@{}c@{}}Primary\\ Outputs\end{tabular}} & \textbf{\begin{tabular}[c]{@{}c@{}}\# Test\\ Pattern\end{tabular}} & \textbf{\begin{tabular}[c]{@{}c@{}}Test Response\\ Length\end{tabular}} & \textbf{\begin{tabular}[c]{@{}c@{}}Compaction\\ Rate\end{tabular}} & \textbf{\begin{tabular}[c]{@{}c@{}}Aliasing\\ Rate\end{tabular}} \\ \hline
\textbf{c17}   & 2                                                                  & 7                                                                  & 14                                                                      & -1728.57$\%$                                                          & 0$\%$                                                            \\
\textbf{c432}  & 7                                                                  & 63                                                                 & 441                                                                     & 41.95$\%$                                                          & 0$\%$                                                            \\
\textbf{c499}  & 32                                                                 & 55                                                                 & 1760                                                                    & 85.45$\%$                                                          & 0$\%$                                                            \\
\textbf{c880}  & 26                                                                 & 148                                                                & 3848                                                                    & 93.35$\%$                                                          & 0$\%$                                                            \\
\textbf{c1355} & 32                                                                 & 100                                                                & 3200                                                                    & 92.00$\%$                                                          & 0$\%$                                                            \\
\textbf{c1908} & 25                                                                 & 128                                                                & 3200                                                                    & 92.00$\%$                                                          & 0$\%$                                                            \\
\textbf{c2670} & 140                                                                & 444                                                                & 62160                                                                   & 99.59$\%$                                                          & 0$\%$                                                            \\
\textbf{c3540} & 22                                                                 & 264                                                                & 5808                                                                    & 95.59$\%$                                                          & 0$\%$                                                            \\
\textbf{c5315} & 123                                                                & 599                                                                & 73677                                                                   & 99.65$\%$                                                          & 0$\%$                                                            \\
\textbf{c6288} & 32                                                                 & 33                                                                 & 1056                                                                    & 75.76$\%$                                                          & 0$\%$                                                            \\
\textbf{c7552} & 108                                                                & 455                                                                & 49140                                                                   & 99.48$\%$                                                          & 0$\%$                                                           
\end{tabular}
}
\label{tab:iscas85_results}
\end{table}  
\section{Conclusion}
\label{sec:conclusion}
In this paper, we tackled the problem of insecure built-in self-test (BIST) relying on a scan chain mechanism. 
Thus, we proposed a (\textsc{KMAC})-based response compaction featuring unique device keys. 
The proposed test method is reliant on (a) a virtual prototype (VP) to generate test golden references, (b) \textsc{KMAC} to generate the signatures of the device under test (DUT), and (c) CPU to schedule the test, increase the DUT availability, and provide a remote testing capability.     
The proposed test method waives the need for a dedicated compaction unit since an existing cryptographic hash engine can be used.
Therefore, sufficiently detailed system-state analysis can be achieved without revealing system internals. 
The compaction process based on \textsc{KMAC} is sufficiently secured against inference due to the use of a unique device key.
We demonstrate the feasibility of our approach using a RISC-V-based SoC.
The results demonstrate that the presented approach could meet all set goals, providing a solid solution for a secure test method.

\bibliographystyle{IEEEtran}
\bibliography{ref}

\end{document}